\documentclass[11pt]{article}
\usepackage{arxiv}
\usepackage{xurl} 
\usepackage{tikz}
\usetikzlibrary{positioning,arrows.meta,fit}
\hypersetup{
  pdftitle={Cascadia: A Control-Plane-Free Alternative to Hyperconverged AI Infrastructure},
  pdfauthor={Matias Parij; Pawan Paudel; Tate Berenbaum; Muthaiah Venkatachalam}
}

\newcommand{\brk}{\discretionary{}{}{}}
\title{Cascadia: A Control-Plane-Free Alternative\\to Hyperconverged AI Infrastructure}

\author{
  \begin{tabular}{@{}c@{\hspace{3em}}c@{}}
    Matias Parij & Pawan Paudel \\
    Not Community Labs Inc. & Not Community Labs Inc. \\[0.75em]
    Tate Berenbaum & Muthaiah Venkatachalam \\
    Not Community Labs Inc. & Intel Corporation
  \end{tabular}
}

\date{September 2026}

\begin{document}

\maketitle

\begin{abstract}
We present Cascadia, a system for serving large language models on fleets of commodity Intel AI PCs using their CPU, integrated-GPU, and NPU resources. Every node embeds ingress, scheduling, and execution; inference requests require no dedicated routing control plane. Nodes join a libp2p QUIC mesh using CA-issued ed25519 admission certificates, gossip signed capabilities, exchange live load over direct peer streams, and route OpenAI-compatible requests to eligible peers. An operator-run certificate authority handles admission and fleet management outside the inference path. Three serving modes share one interface: whole-model execution on one node, load-balanced replicas, and pipeline-sharded chains using the compilation and speculative-decoding mechanism of our companion paper. Optional KV-cache mobility reuses compatible conversation prefixes after a routing move, with cold recomputation on a miss. Signed response receipts and hash-chained logs support provenance and audit. A three-node Phi-3.5-mini NPU testbed delivered $3.10\times$ the response throughput of its one-node configuration under ten concurrent requests; a separate four-node deployment recorded $4.06\times$ the throughput of direct single-node serving. Paired latency observations, runtime measurements, and internal functional checks characterize the tested configurations. We compare Cascadia with IBM, Nutanix, VMware, and HPE platforms on deployment footprint, hardware requirements, scheduling, scaling, licensing, and trust, using vendor documentation. The paper repository provides benchmark scripts, curated measurements, and a claim-to-evidence map.
\end{abstract}

\section{Introduction}
\label{sec:intro}

Organizations now field fleets of AI PCs. Each Intel Core Ultra machine ships an integrated GPU, an NPU, and 16--32\,GB of unified memory, and spends most of its life idle. Treated collectively, such a fleet is a serving platform of meaningful size, but only if software solves problems that datacenter serving stacks were never built for. The machines are heterogeneous and intermittently available; they sit on ordinary office and residential networks behind NAT; no machine can be assumed dedicated to coordination, ingress, or scheduling; trust cannot derive from a cluster boundary, because there is no cluster; and the models worth serving may not fit on any one member. This paper asks what it takes to turn a fleet of client machines into a dependable LLM serving system, and gives a measured answer.

The question is open because the established on-premises architecture answers a different one. Enterprise AI platforms such as IBM's watsonx on OpenShift, Nutanix Enterprise AI on Kubernetes, and their hyperconverged peers assume a dedicated cluster: dedicated control-plane services and server-class hardware, with model execution managed inside that cluster~\cite{rhoaiinstall2026, nutanixnaireq2025, nutanixnaifaq2026, rhelai2025hw}. Those assumptions are sound at their design point and structurally inapplicable to a client fleet; \S\ref{sec:related_incumbent} examines them and \S\ref{sec:comparison} returns with a full comparison. They also leave the fleet's capacity stranded at a time when dedicated AI infrastructure is itself frequently under-utilized~\cite{venturebeat2026gpu}.

In our companion paper~\cite{berenbaum2026shards} we showed that the gap between those two realities is software, not hardware. Pre-compiled pipeline shards with mask-based speculative decoding let a fleet of Intel AI PCs serve models at and beyond the capability of any single member, up to a 70B-parameter model spread across four laptops. That paper established the \emph{mechanism}. This paper presents the \emph{system} that operationalizes it: Cascadia, a distributed serving platform in which every node runs one identical static binary and request ingress, scheduling, certificate verification, and audit are embedded in each node; the CA retains certificate issuance and fleet management.

Cascadia answers the fleet problem with four design decisions. No dedicated request-routing control plane: nodes discover each other via Kademlia, gossip signed capability advertisements, and any node can accept a request, schedule it, or execute it. No dedicated hardware: the serving substrate is the CPU, integrated GPU, and NPU silicon already present in the fleet, driven through OpenVINO. No replica-only ceiling: when a model fits on one node it is replicated and load-balanced; when it does not, it is served as a pipeline-sharded chain across several nodes; when even aggregate RAM is insufficient, a sparse mixture-of-experts engine streams experts from disk. No platform-mediated trust: admission is a CA-signed certificate, every response carries an ed25519 receipt the caller can verify offline, and every node appends to a hash-chained audit log.

We make five contributions:

\begin{enumerate}[leftmargin=*]
  \item \textbf{A control-plane-free architecture for enterprise LLM serving on commodity Intel AI PC fleets}~(\S\ref{sec:overview}, \S\ref{sec:mode_replica}). One binary per node; a libp2p QUIC mesh; gossip-propagated, ed25519-signed capability advertisements and a direct load-state plane; a per-node load-aware scheduler with health-tracked rerouting; and NAT traversal for residential and branch-office nodes. Request scheduling remains local to each node; operator fleet management is off the request path.
  \item \textbf{Unification of three serving modes behind one OpenAI-compatible surface}~(\S\ref{sec:modes}): single-node whole-model, replicated whole-model, and pipeline-sharded chains (including sparse-MoE chains with disk-streamed experts), with a mode-selection analysis and a byte-blind boundary between the orchestration layer and the inference runtime. Session affinity and optional KV transfer manage reusable state when routing changes~(\S\ref{sec:kv_mobility}).
  \item \textbf{A cryptographic trust layer}~(\S\ref{sec:trust}): CA-issued admission certificates, partner and operator PASETO tokens, per-response signed receipts with an external verifier, hash-chained per-node event logs, and mesh-wide revocation measured at 18--20\,s against a $\le$60\,s target.
  \item \textbf{A measured evaluation on real AI PC fleets}~(\S\ref{sec:eval}): $3.10\times$ response throughput at three nodes relative to one in the June study; a separate four-node result of $4.06\times$ over direct single-node serving and $6.86\times$ lower p50 latency under load; and scoped runtime observations for placement and sparse-MoE serving. The evaluation distinguishes performance measurements from functional validation and identifies each measurement session.
  \item \textbf{A scoped comparison with hyperconverged enterprise AI infrastructure}~(\S\ref{sec:comparison}): IBM (watsonx, OpenShift AI, Fusion HCI), Nutanix (Enterprise AI, NKP), VMware (Private AI Foundation), and HPE (Private Cloud AI) on minimum footprint, accelerator gating, control-plane requirements, scaling model, licensing meter, and air-gap posture, sourced exclusively from vendor documentation, with the vendors' counter-arguments stated.
\end{enumerate}

We do not attempt a total-cost-of-ownership model, but the capital-expenditure asymmetry is structural. The incumbent stacks meter licenses against provisioned capacity~\cite{awswatsonx2026, nutanixlicensing2026}, on top of the required server infrastructure. Cascadia runs on hardware the organization already owns, so adding serving capacity does not require renting third-party compute or buying new servers dedicated to inference.

\section{Background and Related Work}
\label{sec:related}

\subsection{The Hyperconverged Enterprise AI Paradigm}
\label{sec:related_incumbent}

The dominant on-premises LLM deployment products share one architectural shape: a managed server cluster with a dedicated coordination layer and provisioned inference hardware.

IBM's stack is watsonx.ai deployed on Red Hat OpenShift via IBM Software Hub (formerly Cloud Pak for Data); serving inference foundation models requires Red Hat OpenShift AI~\cite{ibmswhubrhoai2026}. OpenShift AI's single-model serving is KServe-based~\cite{kserve2026}, and its documented 2.16 serverless configuration additionally requires the OpenShift Serverless (Knative) and Service Mesh (Istio) operators~\cite{rhoaiserving2026}. Accelerator support arrives through vendor operators, with NVIDIA GPUs the primary path and Intel Gaudi~3 a technology preview at the RHEL AI layer~\cite{rhelai2025hw}. IBM also fields turnkey rack appliances (Fusion HCI with NVIDIA L40S and H100~NVL GPU nodes~\cite{ibmfusion2025, ibmfusiongpu2026}; Lenovo's validated design scales to 24 GPUs per rack~\cite{lenovovd2025}) and its own inference silicon for Z and Power systems~\cite{ibmspyre2025}. Inference serving is vLLM-based~\cite{kwon2023vllm} via Red Hat AI Inference Server~\cite{redhataiis2026}, with the Granite model family pre-optimized for it~\cite{granite2025}.

Nutanix Enterprise AI (NAI), the productized successor of GPT-in-a-Box~2.0, deploys as containers on any CNCF-certified Kubernetes (most natively the Nutanix Kubernetes Platform on the NCI/AHV hyperconverged stack) and serves models through NVIDIA NIM microservices or a validated Hugging Face list~\cite{nutanixnaifaq2026, nutanixnvidia2025}. Its documented minimum is three control-plane plus three worker nodes~\cite{nutanixnaireq2025}. A CPU-only mode exists for server-class Xeons with AVX-512 and AMX, which Nutanix positions for models up to roughly 10B parameters~\cite{nutanixamx2025}; the announced NKP Metal targets edge bare metal but remains a Kubernetes cluster on server hardware~\cite{nkpmetal2026}.

VMware's Private AI Foundation with NVIDIA is an add-on to VMware Cloud Foundation: deep-learning VMs and GPU-enabled Kubernetes clusters provisioned through the VCF stack (SDDC Manager, vCenter, NSX, Supervisor) with vLLM and NIM model runtimes layered above. Deployment requires at least three GPU-enabled hosts, and licensing stacks a per-core VCF subscription with per-GPU NVIDIA AI Enterprise; the 9.0 release folded VCF's own private-AI services into the base subscription, while the NVIDIA add-on remains separately licensed~\cite{vmwarepaif2026, vmwarepais2025, broadcomexplore2025}. HPE Private Cloud AI ships the same shape as a turnkey appliance: G1 production rack configurations from four L40S GPUs upward, three dedicated control nodes per rack, and a management plane hosted in HPE's GreenLake cloud, with a G2 air-gapped configuration~\cite{hpepcai2026}.

We return to these systems in \S\ref{sec:comparison}. Their deployment assumptions center on managed server clusters with dedicated coordination services. Our comparison concerns the hardware and operational requirements of those documented configurations. Distributed model execution is also available in datacenter runtimes; it is not unique to Cascadia.

\subsection{Distributed Inference on Commodity Hardware}

Petals~\cite{borzunov2023petals} demonstrated volunteer pipeline inference across consumer NVIDIA GPUs over the internet; Parallax~\cite{lin2024parallax} targets decentralized fleets of Apple Silicon and NVIDIA machines; and MDI-LLM~\cite{wang2025mdillm} studies model-distributed inference on edge devices. exo, in its 2026 form, pools Macs and workstations with automatic discovery, latency- and bandwidth-aware topology, and both pipeline- and tensor-parallel placement, with an Apple-Silicon focus and Linux limited to CPU execution~\cite{exo2024}. On the academic side, prima.cpp~\cite{primacpp2025} and TPI-LLM~\cite{tpillm2024} schedule pipelined-ring and tensor-parallel inference across low-resource home devices. Among open-source servers, LocalAI provides peer-discovered whole-model federation and llama.cpp-RPC workers attached to a serving instance~\cite{localai2026}, and GPUStack manages pooled accelerators under a central server, dropping macOS and native Windows workers in its v2 redesign~\cite{gpustack2026}, an indication of how hard consumer endpoints are to retain. Cascadia shares the premise of this family, that idle consumer silicon is a serving substrate, and differs in target (Intel CPU, iGPU, and NPU via OpenVINO~\cite{openvino2026}), in shipping an enterprise trust layer (admission, receipts, revocation; \S\ref{sec:trust}), in baking no seed, server, or coordinator role into any node, and in unifying replicated whole-model serving with pipeline sharding under one scheduler rather than treating sharding as the only mode.

The Kubernetes-native position on distributed inference is articulated by llm-d, a Red Hat--led framework with Google, IBM, NVIDIA, and others among its founding contributors, that layers KV-cache-aware routing and disaggregated prefill and decode over vLLM on datacenter accelerators~\cite{llmd2025}. NVIDIA NIM documents multi-node tensor- and pipeline-parallel serving on GPU clusters under LeaderWorkerSet or Ray orchestration~\cite{nimmultinode2026}. Datacenter serving systems such as vLLM~\cite{kwon2023vllm}, Sarathi-Serve~\cite{agrawal2024sarathi}, Splitwise~\cite{patel2024splitwise}, and TD-Pipe~\cite{zhang2025tdpipe} optimize GPU-cluster serving. Cascadia's distinction is its client-device substrate and per-node request scheduling.

Closest to Cascadia's substrate are the silicon vendors' own client-device efforts, which stop short of a platform. AMD's first-party guide runs a trillion-parameter MoE across four Ryzen AI Max+ machines with hand-configured llama.cpp RPC, a worker list in a config file, with no discovery, placement, replication, admission, or receipts~\cite{amdcluster2026}; NVIDIA documents clustering DGX Spark units, two over a direct cable and four over a 200\,GbE switch~\cite{nvidiasparkstack2026}. Both validate the workload on client-class silicon while underscoring the absence of a fleet layer. The products that target AI PCs directly are single-device by design: Microsoft's Foundry Local documentation states it ``isn't designed as a server inference stack''~\cite{msfoundrylocal2026}, and LM Studio's enterprise offering governs fleets of single-device installs without pooling them into one endpoint~\cite{lmstudioent2026}. We are not aware of an existing layer that pools Intel client machines into a single served endpoint the way this paper describes.

KV state movement is established in disaggregated serving. DistServe separates prefill and decoding to manage their different resource demands~\cite{zhong2024distserve}; Mooncake organizes serving around a distributed KV cache spanning GPU-cluster resources~\cite{qin2024mooncake}. Cascadia addresses a related placement problem when a conversation moves between pipeline chains in a permissioned client fleet. Its affinity, compatibility checks, restoration, and cold fallback are described in \S\ref{sec:kv_mobility}. We evaluate those mechanisms through scoped functional checks, without importing performance claims from disaggregated datacenter systems.

On the runtime side, Cascadia's sparse-MoE engine adopts techniques from PowerInfer and SmallThinker~\cite{song2024powerinfer, song2025smallthinker} (bounded expert caches, sparsity-aware offload), KTransformers~\cite{chen2025ktransformers} (CPU/GPU expert placement), CHESS-style activation sparsification~\cite{he2024chess}, expert top-$K$ reduction~\cite{fastermoe2025}, and prompt-lookup drafting~\cite{saxena2023pld}.

Following the companion paper's discipline, we benchmark against the strongest same-hardware baseline (direct single-node serving on the identical machine and model) and report absolute numbers, rather than benchmarking against systems whose hardware targets do not overlap ours.

\subsection{The Companion Paper: the Sharded-Chain Mechanism}
\label{sec:related_companion}

This paper treats pipeline-sharded execution largely as a black box whose mechanism is established in the companion paper~\cite{berenbaum2026shards}: per-stage INT4 OpenVINO IR shards exported at monolithic parity (within 0.4\% via a post-export \texttt{beam\_idx} Gather injection); speculative decoding made practical on stateful models by mask-based KV-cache rewind ($1.33\times$ mean single-node, about $1.6\times$ at 2048-token generations); and multi-user micro-batching via independent stateful InferRequests ($1.80\times$ at two streams). Composed, a two-node fleet served two users at $1.79\times$ monolithic single-user throughput; a three-node fleet reached 64.67\,tok/s at three streams; and a 4-stage Llama~3.1~70B INT4 deployment across four Lunar Lake AI PCs reached interactive throughput on a model that fits no single member, validated bit-exact against its same-topology baseline. Those results were measured on the Python research stack; the Rust runtime described here reimplements the same export formats and wire protocol.

\section{System Overview}
\label{sec:overview}

\subsection{Two Planes, Three Binaries}

Cascadia is one system with two planes, an \emph{inference runtime} and a \emph{mesh orchestration plane}, shipped as three binaries. \texttt{cascadia-node} runs on every fleet machine and is identical everywhere: it embeds the HTTPS gateway, the scheduler, the gossip and discovery subsystems, the trust machinery, and the inference backends. \texttt{cascadia-ca} is an operator-run certificate authority that issues admission certificates and partner tokens; it also exposes fleet inventory and deployment jobs. It sits off the inference request path, and its root key never links into node binaries. \texttt{cascadia-relay} is a Circuit Relay~v2 service for NAT-restricted peers. The runtime is a separate Apache-2.0 Rust workspace consumed by \texttt{cascadia-node} as a library; its \texttt{Engine} and \texttt{Builder} traits are the plugin seam behind which the engines sit.

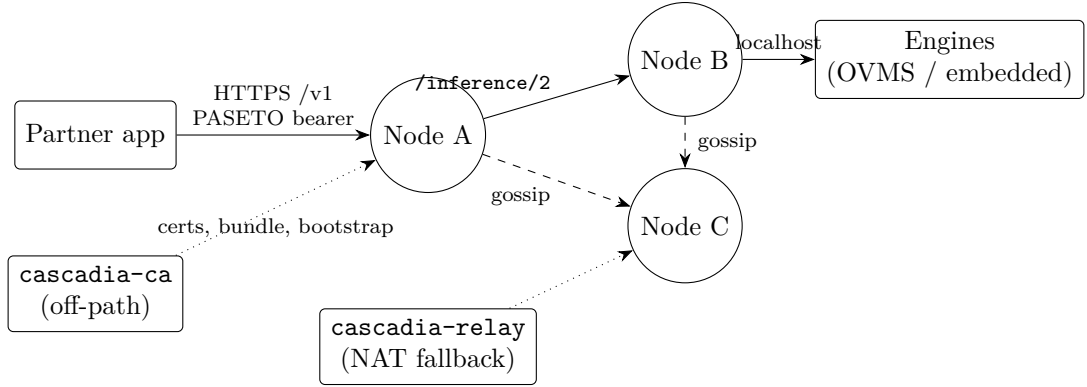
\begin{figure}[t]
\centering
\begin{tikzpicture}[
  node/.style={circle, draw, minimum size=1.5cm, font=\small, align=center},
  svc/.style={rectangle, draw, rounded corners=2pt, minimum height=0.9cm, font=\small, align=center, inner sep=4pt},
  lbl/.style={font=\scriptsize, align=center},
  arr/.style={-{Stealth[length=2mm]}},
]
\node[svc] (partner) at (-4.4, 1.2) {Partner app};
\node[node] (a) at (0, 1.2) {Node A};
\node[node] (b) at (3.4, 2.2) {Node B};
\node[node] (c) at (3.4, 0.0) {Node C};
\node[svc] (ovms) at (6.9, 2.2) {Engines\\(OVMS / embedded)};
\node[svc] (ca) at (-4.4, -0.9) {\texttt{cascadia-ca}\\(off-path)};
\node[svc] (relay) at (0, -1.6) {\texttt{cascadia-relay}\\(NAT fallback)};
\draw[arr] (partner) -- node[lbl, above] {HTTPS /v1\\PASETO bearer} (a);
\draw[arr] (a) -- node[lbl, above left=-2pt] {\texttt{/inference/2}} (b);
\draw[arr, dashed] (a) -- node[lbl, below left=-2pt] {gossip} (c);
\draw[arr, dashed] (b) -- node[lbl, right=1pt] {gossip} (c);
\draw[arr] (b) -- node[lbl, above] {localhost} (ovms);
\draw[arr, dotted] (ca) -- node[lbl, below] {certs, bundle, bootstrap} (a);
\draw[arr, dotted] (relay) -- (c);
\end{tikzpicture}
\caption{One binary, every node identical. A partner request enters at \emph{any} node; that node authenticates the bearer token, picks the best peer by capability and load (possibly itself), forwards over an authenticated libp2p QUIC stream if needed, and streams tokens back wrapped in a signed receipt. The CA is off the inference request path; the relay carries peer traffic when direct connectivity is unavailable.}
\label{fig:mesh}
\end{figure}

There is no gateway tier, no scheduler tier, no worker tier. Any node can accept a partner request, schedule it, execute it locally, or forward it to a peer (Figure~\ref{fig:mesh}). We use \emph{control-plane-free} to mean that inference requests require no dedicated routing or scheduling service. Admission, revocation publication, and operator-initiated fleet changes still depend on management services: the CA now resolves deployment intents into per-node commands and tracks their outcomes. This management role is distinct from per-request scheduling.

\subsection{Design Tenets}
\label{sec:tenets}

Five decisions, each later contrasted with the incumbent stacks in \S\ref{sec:comparison}:

\begin{enumerate}[leftmargin=*]
  \item \textbf{No dedicated tiers.} Entry, scheduling, and execution are roles every node can play per request, not machine types. Request routing has no dedicated coordinator; losing a stage can still interrupt its chain.
  \item \textbf{OpenAI-compatible at every node.} The partner surface is \texttt{/v1/chat/completions} and \texttt{/v1/models} over HTTPS with bearer auth, so existing SDKs work unmodified. \texttt{/v1/models} aggregates across the mesh, so one endpoint exposes every model the fleet serves.
  \item \textbf{Trust via cryptography, not platform boundary.} Admission certificates, signed capability advertisements, per-response receipts, and hash-chained event logs make trust claims independently checkable rather than implied by cluster membership~(\S\ref{sec:trust}).
  \item \textbf{Engines are pluggable; the orchestration layer is byte-blind.} The mesh moves activation bytes for sharded chains without interpreting them, and tensor codecs belong to the runtime~(\S\ref{sec:chains}). New engines arrive behind the \texttt{Engine} and \texttt{Builder} traits without touching the mesh.
  \item \textbf{Intel-native, single static binary.} Rust throughout; OpenVINO reaches CPU, integrated GPU, and NPU on Meteor, Lunar, Arrow, and Panther Lake AI PCs. Node releases and installers now cover Windows and Ubuntu Linux; the CA and relay ship Linux releases. Device support still depends on the platform and driver. There is no Python, container runtime, or cluster install on workers.
\end{enumerate}

\subsection{Request Lifecycle}
\label{sec:lifecycle}

A partner sends \texttt{POST /v1/chat/completions} with a PASETO v4.public bearer token to any node. The receiving node verifies the token (signature, expiry, revocation cache), then its scheduler picks the serving peer by capability, load, and peer-health rank, possibly itself, which collapses the forwarding hop entirely. A forwarded request travels as length-prefixed bincode frames over \texttt{/cascadia/inference/2}, a libp2p stream inside the QUIC session (TLS~1.3 on direct links, Noise on relayed circuits); the worker deduplicates request IDs for replay defense, executes against its local engine, and streams a \texttt{Metadata} frame (announcing \texttt{served\_by}), then \texttt{Chunk} frames, then a terminal signed \texttt{Receipt} frame. The entry node enforces an eligibility gate on \texttt{served\_by} (a response from an unadmitted or revoked peer is rejected with 502), translates frames to SSE, and places the receipt in the final data event before \texttt{[DONE]}. First-token and inter-token deadlines convert a stalled peer into a typed error response rather than a hung connection, and a failed first peer is retried once on an alternate before surfacing 503. Both sides append the exchange to their hash-chained event logs.

The implementation described here uses \path{/cascadia/inference/2}, \path{/cascadia/gossip/5}, \path{/cascadia/state/v2}, and \path{/cascadia/pipeline/1}; KV mobility uses a separate \path{/cascadia/state/kv/v1} stream. Each protocol is versioned independently; a breaking change requires compatible peers to upgrade. The two system planes above describe the runtime/mesh boundary; the load-state and KV coordination planes are protocols inside the mesh.

\section{Serving Modes}
\label{sec:modes}

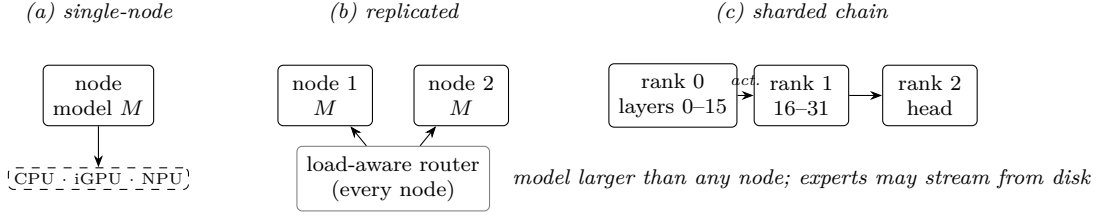
\begin{figure}[t]
\centering
\begin{tikzpicture}[
  box/.style={rectangle, draw, rounded corners=2pt, minimum width=1.25cm, minimum height=0.8cm, font=\scriptsize, align=center},
  dev/.style={rectangle, draw, dashed, rounded corners=2pt, font=\tiny, align=center, inner sep=2pt},
  lbl/.style={font=\scriptsize\itshape, align=center},
  arr/.style={-{Stealth[length=1.6mm]}},
]
\node[lbl] at (0, 2.1) {(a) single-node};
\node[box] (m1) at (0, 1.0) {node\\model $M$};
\node[dev] (m1d) at (0, -0.1) {CPU $\cdot$ iGPU $\cdot$ NPU};
\draw[arr] (m1) -- (m1d);
\node[lbl] at (3.9, 2.1) {(b) replicated};
\node[box] (m2a) at (3.0, 1.0) {node 1\\$M$};
\node[box] (m2b) at (4.8, 1.0) {node 2\\$M$};
\node[box, draw=gray] (m2s) at (3.9, -0.1) {load-aware router\\(every node)};
\draw[arr] (m2s) -- (m2a);
\draw[arr] (m2s) -- (m2b);
\node[lbl] at (9.3, 2.1) {(c) sharded chain};
\node[box] (r0) at (7.6, 1.0) {rank 0\\layers 0--15};
\node[box] (r1) at (9.3, 1.0) {rank 1\\16--31};
\node[box] (r2) at (11.0, 1.0) {rank 2\\head};
\draw[arr] (r0) -- node[lbl, above=0pt] {\tiny act.} (r1);
\draw[arr] (r1) -- (r2);
\node[lbl] at (9.3, -0.1) {model larger than any node; experts may stream from disk};
\end{tikzpicture}
\caption{Three serving modes behind one OpenAI-compatible surface. A fleet runs all three at once; mode choice is per model.}
\label{fig:modes}
\end{figure}

The same mesh serves models in three modes (Figure~\ref{fig:modes}), selected per model at deployment time. The scheduler routes per request across all three: whole-model requests to the least-loaded replica, sharded requests to the entry rank of the least-loaded feasible chain.

\subsection{Mode 1: Single-Node Whole-Model}
\label{sec:mode_single}

A model that fits one node's memory runs behind one of two backend families. The \emph{OVMS backend} supervises an OpenVINO Model Server subprocess (batching, KV-cache management, NPU offload, and an OpenAI surface arrive for free) with crash-loop detection and backoff; this is the production path for NPU-resident INT4 models such as Phi-3.5-mini and Llama~3.1~8B in our fleets. The CPU/GPU OVMS path exposes continuous-batching and prefix-cache settings; its node admission limit must be set high enough for concurrent requests to reach the batcher. NPU configurations reject those batching options and remain sequential. The \emph{embedded engines} link the runtime directly into \texttt{cascadia-node}: \texttt{ov-genai} wraps \texttt{openvino\_genai} pipelines and adds speculative decoding via a FastDraft-class draft model or prompt-lookup drafting~\cite{saxena2023pld}. On Llama~3.1~8B INT4 on an Arc B390 integrated GPU, 20.83\,tok/s plain rises to 28.04\,tok/s with a 150M draft at $K{=}5$ (runtime-repo measurement, May~2026).

Within a node, placement across CPU, iGPU, and NPU is itself an optimization problem because all three share one unified-memory pool. The runtime profiles per-device compile time, latency, and operator support, then solves an exact branch-and-bound ILP (adapted from PowerInfer's formulation~\cite{song2024powerinfer} to Intel's UMA constraint) that assigns stages to devices under per-device and global memory caps. This placement path is merged to the runtime's main branch. In the cases measured so far (a four-trial mean with $\sigma<2\%$ on the headline workload), it pays off in the memory-pressure regime: a Yi-1.5-9B fp16 workload at 99.8\% of nominal iGPU capacity improves from 2.03 to 2.91\,tok/s ($+43\%$) by moving the embedding to CPU. Comfortably-fitting models are best left on the iGPU alone, and the NPU was the slowest decode tier in every configuration we measured~(\S\ref{sec:eval_runtime}).

\subsection{Mode 2: Replicated Whole-Model}
\label{sec:mode_replica}

When several nodes hold the same model, Cascadia provides replica scaling without a dedicated cluster. Each node gossips an ed25519-signed \texttt{CapabilityAd} describing its models, engines, shards, and hardware. Live load no longer rides in that advertisement: direct, transport-authenticated \texttt{LoadFrame}s populate a separate \texttt{LoadCache}, while signed capabilities remain in the TTL-evicted, epoch-replay-protected \texttt{MeshView}. The scheduler combines these views with local in-flight accounting and peer health. A dynamic capacity envelope publishes request occupancy, memory availability, and headroom; local admission reservations enforce concurrency limits, so a ranking decision alone cannot over-admit work. A \texttt{PeerHealthTracker} demotes failing peers and requires recovery evidence before restoring their rank. First-peer failures are retried on an alternate peer before surfacing a typed error. On reconnect, the state stream also re-sends signed capability and chain advertisements, avoiding reliance on a fresh gossip flood for chain re-formation.

Each node can expose a multi-model fleet through one entry point: \texttt{/v1/models} is the mesh union. A heterogeneous mesh routes each request using capability and load; we have run multi-model mixes of Phi-3.5-mini with Qwen3-8B and with Llama~3.1~8B~(\S\ref{sec:eval_method}).

\subsection{Mode 3: Pipeline-Sharded Chains}
\label{sec:chains}

A model larger than any node's memory is exported once into per-stage shards (the companion paper's INT4 OpenVINO IR format) and served by an ordered chain of workers. The orchestration plane treats chains as a routing problem: workers gossip \texttt{ShardDescriptor}s for the stages they hold, a \texttt{Chain\brk Assignment\brk Ad} binds an ordered set of workers into a serving chain, and admission of a chain is gated by feasibility rules covering contiguity and completeness of stages, per-worker memory fit, engine-kind agreement, and engine-build-version homogeneity. Activation traffic flows over \texttt{/cascadia/pipeline/1}, which the mesh treats as an opaque byte stream bridged onto the runtime's loopback TCP relay. The orchestration plane is byte-blind, so the tensor wire format (a 20-byte header, f16 hidden states, 8\,KB per hop at 8B scale~\cite{berenbaum2026shards}) belongs to the runtime alone and the two layers version independently. Rank~0 holds the tokenizer and drives the generation loop; same-node chain edges collapse to in-process duplex streams. When several workers hold the same stage ranks, the scheduler enumerates every feasible chain rather than a single one, pairing the per-rank holders into parallel chains, and the entry node balances requests across them by in-flight load. A sharded model therefore scales horizontally by adding chains, as a whole-model deployment scales by adding replicas.

The chain engines include the following. \texttt{ov-runtime} drives stateful per-stage shards, and through a host-side static KV ring it also drives the stateless exports the NPU requires. \texttt{ov-dist-spec} adds distributed speculative decoding: the driver holds a draft model (optionally on a different device than stage~0) and verifies $K$-token drafts through the chain in single round trips, with the mask-based KV rewind of the companion paper applied driver-side. On a 2-stage Thunderbolt-linked pair this reached 29.62\,tok/s at 4096-token generations (runtime measurement, May~2026), and overlapping network and draft compute adds 6--19\% depending on $K$. A Gemma-4-specific engine handles that family's cross-stage shared-KV forwarding, where KV tensors produced in one stage are consumed by layers in a later stage and cross the wire as tagged frames.

The companion paper reports a four-stage Llama~3.1~70B INT4 chain across four AI PCs on its research runtime~\cite{berenbaum2026shards}. The mesh evidence described here is distinct: internal hardware checks exercise a four-stage Llama~3.1~8B chain through the encrypted pipeline bridge, including session routing and explicit failure reporting. A mixed-serving configuration combines two NPU Phi-3.5 replicas with a two-stage GPU Llama chain behind one entry. These are functional validations of composition and routing. The raw-runtime 70B result is not a measurement of the encrypted mesh path. Cross-chain KV checks establish the additional state-transfer behavior in \S\ref{sec:eval_kv}.

\subsection{Mode 3, Extreme Case: Sparse-MoE with Disk-Streamed Experts}
\label{sec:moe}

Sparse mixture-of-experts models break the assumption that serving capacity is bounded by accelerator memory, or even by aggregate memory, because per token only a few experts in each layer activate. Cascadia's sparse-MoE engine serves Kimi K2.6 (61 layers, 384 experts per MoE layer, top-8 sigmoid routing, MLA attention, roughly 553\,GB of INT4 weights) on hosts with a fraction of that RAM. It memory-maps experts directly from the model's safetensors shards and dispatches them through a purpose-built AVX-512 INT4 GEMM kernel written for this access pattern. A bounded LRU expert cache and a same-as-last-token prefetcher (in the lineage of PowerInfer and SmallThinker~\cite{song2024powerinfer, song2025smallthinker}) keep hot experts resident; attention shells and the embedding run through pure-Rust INT4 and bf16 kernels.

Two sparsity controls trade computation against approximation, in line with recent MoE-reduction results~\cite{fastermoe2025, he2024chess}. The first dispatches only the top-$K'$ of the router's eight experts; the $K'{=}4$ configuration in Table~\ref{tab:moe} has higher throughput than the router-faithful configuration, but changes the model's computation and outputs. The second applies CHESS-style per-channel FFN activation thresholds with an AXPY-form down-projection kernel. Speculative decoding uses a zero-compute n-gram draft~\cite{saxena2023pld} with a multi-token tiled verify that amortizes expert loads across draft positions. For pipeline-parallel chains, a \texttt{ForwardBatch} frame carries $K$ verify positions in one round trip, replacing $K$ separate exchanges. The numerical MoE observations in this paper concern single-host execution.

On the one host measured to date (a Xeon Gold 6252 with 133\,GB of RAM, where the model does not fit in memory and serving is disk-bound), K2.6 produces 0.105 output tokens per second at $K'{=}8$, rising to 0.325\,tok/s at $K'{=}4$~(\S\ref{sec:eval_runtime}). This is a capability demonstration rather than an interactive-serving claim. Its significance is architectural: the ceiling becomes storage bandwidth rather than accelerator memory, and the same chain machinery of \S\ref{sec:chains} can distribute MoE layers across fleet members.

\subsection{Mode Selection}
\label{sec:mode_selection}

The decision boundary is primarily memory and secondarily workload shape. If a model fits one node with headroom, replicate it (mode~2): per-stage splitting costs roughly 11--15\% overall relative to the unsplit model~\cite{berenbaum2026shards}, so sharding a model that fits adds overhead in the cited single-stream configuration, though chains recover and surpass parity at multi-stream operating points through micro-batching and speculative decoding ($1.79\times$ monolithic single-user at two streams in the companion measurements). If a model exceeds one node but fits the fleet's aggregate accelerator memory, serve it as a chain (mode~3); if it exceeds even that, the sparse-MoE path streams experts from disk~(\S\ref{sec:moe}). The modes coexist on one fleet: the mixed rig of \S\ref{sec:chains} serves NPU-resident whole-model replicas and a GPU sharded chain behind the same entry at the same time.

\subsection{Session Affinity and KV-Cache Mobility}
\label{sec:kv_mobility}

Balancing requests across chains creates a state-placement problem: a follow-up turn can arrive at a chain that has the model but lacks the conversation's cached keys and values. Cascadia first avoids unnecessary moves. A tenant-namespaced session hint, supplied through \texttt{X-Cascadia-Session} or derived from the conversation, indexes an entry-local affinity map with bounded retention. The scheduler prefers the previous chain while it remains eligible and below a load ceiling; persistent overload or loss makes the session movable. Affinity is a routing hint, not a durable session directory: losing it can cause cold recomputation.

\textbf{Pull on a forced move.} The KV coordination extension adds \texttt{/cascadia/state/kv/v1} and an engine export/import interface. The entry supplies the previous chain as a hint to the destination head. After rendering and tokenizing the new turn, that head negotiates a reusable prefix with a prior holder. The offer identifies a snapshot epoch and prefix length; the consumer fetches the participating ranks' state and checks the model fingerprint, engine revision, KV layout, token-prefix identity, payload sizes, and checksums before insertion. The runtime owns the snapshot representation, including opaque engine state; the mesh transports it without interpreting tensors. A shared snapshot holder answers export requests without taking the generation engine's lock, allowing a busy source to serve the pull.

\textbf{Consistent restoration.} In the default restoration path, the destination head gathers the source ranks' snapshots and carries downstream state through the pipeline restore path. A separately gated path lets destination ranks fetch their own slices, stage them, and confirm readiness before the head commits the restore; abort handling prevents a failed preparation from authorizing a partial warm start. These paths must agree on the snapshot epoch and prefix depth. A hit counter alone does not establish successful restoration: the resumed turn must actually consume the imported state on the required ranks (\S\ref{sec:eval_kv}).

\textbf{Bounded fallback and rollout.} A minimum reusable-prefix threshold, pull deadline, transfer-size limits, and in-flight reservations bound speculative work. A missing holder, incompatible snapshot, resource refusal, or timeout sends the request to cold prefill. Cold fallback preserves the ability to recompute the request; a failed pull can still add waiting time, and a successful transfer is not necessarily faster than prefill. The relevant comparison is negotiation plus transfer and restore time against the prefill time avoided, which depends on the model, context, and link. Release builds include \texttt{kv\_coord}, but its runtime switch is off by default and enabling it also requires a model allowlist (or an explicit allow-all override). Bounded replica storage and replica lookup are implemented, but the cross-chain checks described in \S\ref{sec:eval_kv} do not establish production durability of replicated KV.

KV mobility concerns reuse of a completed conversation prefix after a routing move. Continuing an already-open response after a chain dies is a separate, also opt-in mechanism: the scheduler can send the token IDs already emitted to an alternate chain as a forced prefix and splice its continuation into the same client stream. This mechanism does not require the failed chain's KV to remain reachable. Unsupported engines or missing trusted token IDs decline resumption and retain explicit failure reporting. Neither mechanism makes arbitrary engines, layouts, or model versions interchangeable.

\section{Trust, Admission, and Provenance}
\label{sec:trust}

The incumbent platforms' trust story is the platform boundary: workloads inside the cluster are trusted because the cluster admitted them, and audit is a logging product. Cascadia's nodes are laptops on office and residential networks, so its trust story is cryptographic and per-artifact, designed so that every claim a partner or operator relies on is independently checkable.

\textbf{Admission.} Each node holds an ed25519 keypair that doubles as its libp2p peer identity. Joining the mesh requires an \texttt{AdmissionCert} signed by the CA root: a bincode-canonical envelope signed raw, intentionally not a bearer-token format. Byte-stable encoding lets the CA deterministically re-derive and re-serve certificates, which enables a polling enrollment flow in which the node generates its key locally, the operator approves the printed public key out of band, and the node fetches its certificate once it is issued. Peers verify each other on \texttt{/cascadia/admission/1} before any inference traffic flows.

\textbf{Three auth roles.} Partners bear PASETO v4.public tokens~\cite{paseto2026}; operators bear separate PASETO operator tokens for CA admin endpoints; nodes bear admission certificates. The CA root anchors admission certificates and partner tokens, while operator tokens are signed by a separate operator root. PASETO's algorithm-locked profile avoids the JWT downgrade-and-confusion class of failures on the bearer paths.

\textbf{Per-response receipts.} The executing node signs a \texttt{Response\brk Receipt} binding the request hash, the response hash, and the executor identity, returned in a \texttt{cascadia} envelope field on both streaming and non-streaming responses. A standalone verifier crate (\texttt{cascadia-verify}) re-derives the hashes and checks the signature offline; cross-language test vectors pin the canonical byte encoding so partner SDKs in any language can verify without linking Rust. For an optionally resumed stream, the scheduler forwards the alternate chain's receipt; it does not construct a multi-executor proof of the interrupted execution. The reserved \texttt{attestation} field is excluded from the receipt's hash domain; a signed receipt is not a hardware-attestation claim.

\textbf{Hash-chained audit logs.} Every node appends signed events (request received, forwarded, served, admission changes) to an append-only log with per-entry \texttt{prev\_hash} chaining, per-entry fsync, and a crash-safe 12-step rotation state machine; \texttt{cascadia-node events verify-chain} fails on any byte flip. The receipts identify the executor and response; the log links each local event to its predecessor.

\textbf{Revocation.} The CA publishes signed revocation bundles, and nodes re-gossip each new bundle on a version change, diffing it locally against the last they saw. Three internal admission trials recorded mesh-wide propagation in 18--20\,s against a $\le$60\,s objective. Applying a revocation closes the affected peer connections. The entry node also gates forwarded responses on its local view of the executor's eligibility (\emph{admitted} $\wedge$ $\neg$\emph{revoked}); rejection therefore depends on the entry learning the revocation. Replay of captured request IDs is rejected worker-side by a TTL'd dedup cache with a deterministic error, and no receipt is re-issued for a rejected replay.

\textbf{KV tenant boundary.} Prefix caches are namespaced by the tenant authenticated at the gateway, and imported entries use the consumer's asserted tenant rather than trusting the snapshot's tenant field. Internal validation includes cross-tenant denial with a same-tenant positive control. This is not isolation from a compromised admitted worker: peer admission authenticates the node, but the KV protocol still trusts tenant assertions relayed by that node. The trust model assumes admitted workers relay tenant identities honestly and restricts KV coordination and replica mutation to that operator-controlled domain. Per-response receipts attest output provenance; they do not prove that imported KV was computed correctly.

\textbf{A constraint from the hardware.} NPU execution is not bit-deterministic across machines, so cross-executor response hashes are not byte-comparable on real NPUs. Receipt verification (that this executor produced this response for this request) is unaffected, and bytewise output equality is not an NPU correctness oracle across executors.

\section{Operability and Resilience}
\label{sec:ops}

A fleet platform lives or dies on day-2 operations performed by people who are not Kubernetes administrators.

\textbf{Setup and packaging.} The \texttt{setup} subcommand on each binary is a one-command interactive wizard; \texttt{doctor} preflights connectivity, configuration, port bindings, and the OVMS toolchain; and \texttt{bootstrap} installs dependencies and fetches models from Hugging Face. Nodes install as a Windows Service or a Linux systemd service. The Windows service uses SCM failure actions (auto-restart on a 5\,s failure-action delay after a forced kill, listener rebind tolerant of TIME\_WAIT); the CA and relay ship systemd units with a hardening profile (privilege drop, namespace and write-exec restrictions) and Docker Compose for overlay deployments. Release CI publishes checksum manifests, and provisioned onboarding scripts install dependencies, enroll the node, and check serving readiness on Windows and Ubuntu. The runtime release workflow builds against pinned OpenVINO GenAI SDK archives and bundles the corresponding libraries. Its hardware tests use the SDK configured on each runner; compatibility is scoped to the tested engine, SDK, driver, device, and model export.

\textbf{NAT traversal.} Residential and branch nodes traverse NATs via AutoNAT~v2 detection, DCUtR hole-punching, and Circuit Relay~v2 fallback~\cite{libp2p2026}, plus UPnP where available; CA-side \texttt{/heartbeat} and \texttt{/bootstrap-peers} endpoints provide a rendezvous by publishing and fetching peer addresses. Heartbeats also carry management telemetry and queued operator commands; they do not choose the executor for an inference request. The functional evidence covers simulated full-cone and symmetric NATs; its scope is those controlled network configurations.

\textbf{Failure behavior.} Before response streaming begins, bounded retries and peer-health demotion route around failures. A dead worker cannot hand off its in-memory KV state; the KV pull mechanism of \S\ref{sec:kv_mobility} requires the source to be reachable, so worker death causes the pull to fail. Without a compatible cached prefix available elsewhere, an eligible retry recomputes the prefix on a replacement worker or complete chain; surviving ranks alone cannot resume an interrupted sharded generation. After an SSE response has started, an exhausted chain emits a typed error event followed by \texttt{[DONE]}, rather than silently presenting a truncated completion as success. Optional forced-prefix resumption attempts continuation on another chain (\S\ref{sec:kv_mobility}); its runtime switch remains off by default. State-plane re-synchronization helps restarted peers rejoin chains, but does not itself preserve an interrupted generation. The functional checks distinguish successful continuation from an explicit terminal failure; they do not imply a recovery-time guarantee.

\textbf{Fleet management.} The CA exposes fleet inventory and operator jobs for deployment, unloading, and stage reassignment. Placement can use reported capacity or an explicit device list, and a job tracks per-device outcomes, including partial failure. In the current heartbeat path, a deployment command records an assignment for the next start; that acknowledgement is not proof that a new sharded model is already serving. These management operations are separate from decentralized request routing.

\textbf{Observability.} Prometheus metrics cover request routing, peer state, admission, inference, event logs, and lifecycle operations; metric-label cardinality is checked by compile-fail tests. KV metrics distinguish pull outcomes and record transferred bytes and latency, while forced-prefix rescue has a separate event. OpenTelemetry traces expose forwarded hops. Metrics bind to localhost by default.

\textbf{Air-gap posture.} A Cascadia deployment is static binaries plus model files: no image registry, no operator catalog, no cluster bootstrap. Disconnected operation is the normal case rather than a separate install path, and the CA runs wherever the operator's trust boundary is.

\section{Evaluation}
\label{sec:eval}

The system description reflects the September 2026 implementation. The performance evaluation combines dated May and June deployments, while chain and KV checks establish functional behavior on their specified configurations. The enterprise mesh is closed source. The \texttt{reproduction/\brk CLAIMS.md} file distinguishes curated public measurements, internal author records, and results cited from the companion paper.

\subsection{Method and Testbeds}
\label{sec:eval_method}

The paired latency and scaling benchmarks compare the same model and hardware class, discard warmups, and report medians and measurement dates. The June latency study alternated requests between direct OVMS and the mesh, with 20 measured requests per arm at concurrency~1. The June scaling study used 60 requests per fleet size at concurrency~10. The released legacy request records capture content arrival, timing, executor identity, and receipt presence. Their success flag denotes a nonempty response, not an independently checked SSE termination sequence; the response-rate observations below use that definition. Every measured mesh response in those released records also carries a receipt. Receipt presence and cryptographic verification are separate properties. The public package includes the records, derived summaries, and the selection of scaling runs.

The May mesh testbed comprised four Intel Core Ultra AI PCs running Phi-3.5-mini-instruct INT4 on NPUs through OVMS, with an identical standalone-OVMS machine as the direct baseline. The June study used Lunar Lake AI PCs with the same model/backend family: a paired two-machine latency setup and a one-to-three-node serving study. The entry can itself serve requests, so the three-node configuration consists of an entry/worker and two additional workers. The May results are retained as a separate historical deployment rather than pooled with June observations. Runtime measurements in \S\ref{sec:eval_runtime} disclose their own machines and trial counts.

\subsection{Observed Latency and Fleet Throughput}
\label{sec:eval_overhead}

\begin{table}[t]
\centering
\small
\begin{tabular}{lrrl}
\toprule
Metric & Direct OVMS & Cascadia 4-node & Result \\
\midrule
TTFT p50, single request & 1438\,ms & 1307\,ms & observed medians \\
Aggregate throughput (10 conc.) & 0.104\,req/s & 0.422\,req/s & $4.06\times$ \\
Wall time, 100 requests & 16\,min & 4\,min & $4.04\times$ \\
p50 latency under load & 87.2\,s & 12.7\,s & $6.86\times$ lower \\
Reported success (600+ requests) & 100\% & 100\% & internal benchmark \\
\bottomrule
\end{tabular}
\caption{Historical four-node mesh versus direct single-node OVMS: Phi-3.5-mini-instruct INT4 on Intel AI PC NPUs, measured 2026-05-13 on the pre-embedded-engine deployment. Source: the authors' internal benchmark summary; the aggregate and loaded-latency observations use ten concurrent requests. The 100-request durations are rounded, and ratios use the original unrounded durations. The reported 600+-request success window is an operational observation, not an interrupted-stream recovery rate. These medians describe the tested configuration and do not constitute a statistical equivalence test.}
\label{tab:overhead}
\end{table}

Table~\ref{tab:overhead} records $4.06\times$ aggregate throughput and $6.86\times$ lower median latency under load for the four-node deployment relative to one direct-serving machine. The authors' benchmark summary also records 22--28\% of requests routed to each worker. A separate paired session on 2026-06-08 observed TTFT p50 of 1286\,ms through the mesh and 1568\,ms through direct OVMS, with 20 measured requests per arm. All 20 mesh responses carried receipts. The paired observations show lower median TTFT for the tested mesh path; they do not isolate the cost of each routing or cryptographic operation. The public records support this latency comparison independently of the May aggregate result.

\subsection{Replica Scaling}
\label{sec:eval_scaling}

Table~\ref{tab:scaling} tracks the replica-scaling curve.

\begin{table}[t]
\centering
\small
\begin{tabular}{rrrrl}
\toprule
Nodes & Concurrency & Responses/s & p50 under load & Relative rate \\
\midrule
1 & 10 & 0.222 & 45.0\,s & $1.00\times$ \\
2 & 10 & 0.484 & 18.9\,s & $2.17\times$ \\
3 & 10 & 0.690 & 12.2\,s & $3.10\times$ \\
\bottomrule
\end{tabular}
\caption{Replica serving measured 2026-06-08 on Lunar Lake AI PCs with Phi-3.5-mini INT4/NPU; 60 requests per point at concurrency~10. The measured response rate rises by $2.17\times$ at two nodes and $3.10\times$ at three, relative to the one-node run. Worker routing counts were 60; 28/32; and 19/19/22, respectively. A run labeled one node that actually routed to two workers is excluded; its records and exclusion rationale are retained with the dataset. The separate four-node result in Table~\ref{tab:overhead} uses a different deployment and session.}
\label{tab:scaling}
\end{table}

\subsection{Trust-Layer Latencies}
\label{sec:eval_trust}

Table~\ref{tab:trust} summarizes the trust-layer timings.

\begin{table}[t]
\centering
\footnotesize
\setlength{\tabcolsep}{4pt}
\begin{tabular}{llll}
\toprule
Quantity & Target & Measured & Source \\
\midrule
Revocation propagation, mesh-wide & $\le$60\,s & 18--20\,s & internal admission trials, Apr 2026 \\
Fourth node reaches mesh-ready (peers known) & n/a & $\le$14\,s & internal mesh-join observation \\
Receipt presence (stream / non-stream) & n/a & 100\% / 100\% & \texttt{e7}, 2026-06-08 \\
Receipt length (base64); \texttt{/v1/models} p50 & n/a & 272 chars; 137\,ms & \texttt{e7}, 2026-06-08 \\
\bottomrule
\end{tabular}
\caption{Trust-layer observations. Revocation propagation comes from three internal admission trials; mesh-ready denotes discovery of the fourth node's peers, not model readiness. The public receipt session was measured 2026-06-08 on a Lunar Lake/Phi-3.5-mini INT4/NPU deployment: 20 streamed responses, ten non-streamed responses, and 50 model-list probes. Each response carried a receipt, of length 272 base64 characters for the streamed responses. The 137\,ms model-list value is a client-observed median. Receipt presence does not measure verification cost.}
\label{tab:trust}
\end{table}

\subsection{Runtime-Level Results}
\label{sec:eval_runtime}

Two engine-layer results (Tables~\ref{tab:placement} and~\ref{tab:moe}) round out the system picture. These runtime-repository measurements (May 2026) are presented as observations rather than headline claims; the placement caption distinguishes its repeated headline workload from the single-trial rows.

\begin{table}[t]
\centering
\footnotesize
\begin{tabular}{lrrl}
\toprule
Workload (Lunar Lake, 32\,GB UMA) & iGPU-only & ILP placement & Note \\
\midrule
Yi-1.5-9B fp16 (16.45/16.48\,GiB iGPU) & 2.03\,tok/s & 2.91\,tok/s & $+43\%$; embedding to CPU \\
Qwen2.5-32B INT4 (comfortable fit) & 1.97\,tok/s & 1.27\,tok/s & iGPU-only wins; ILP overfits \\
SOLAR-10.7B fp16 (over-nominal) & 1.31\,tok/s & 0.79\,tok/s & transparent UMA spill unmodeled \\
\bottomrule
\end{tabular}
\caption{Intra-node CPU/iGPU/NPU placement (Yi-1.5-9B is a four-trial mean, $\sigma<2\%$; the other rows single-trial; May 2026; merged to the runtime main branch). The ILP pays off only in the near-full-iGPU pressure regime; we report the cases it loses because the boundary is the finding. NPU-overflow placements were strictly worse in every configuration measured.}
\label{tab:placement}
\end{table}

\begin{table}[t]
\centering
\small
\begin{tabular}{lrr}
\toprule
K2.6 expert dispatch & tok/s & vs.\ $K'{=}8$ \\
\midrule
Top-$K'{=}8$ (router-faithful) & 0.105 & baseline \\
Top-$K'{=}4$ & 0.325 & $+210\%$ \\
\bottomrule
\end{tabular}
\caption{Sparse-MoE serving of Kimi K2.6 ($\sim$553\,GB INT4) on one disk-bound Xeon Gold 6252 host with 133\,GB RAM. The May 2026 internal runtime experiment used greedy decoding with up to 64 generated tokens per prompt and reported one session per configuration. The reported output-token rate includes request wall time. The model is about $4\times$ larger than host RAM. A separate session recorded 0.156\,tok/s at $K'{=}6$ versus 0.115\,tok/s at $K'{=}8$ ($+36\%$); it is not pooled with the table's session. Expert reduction changes the model's computation, so these throughput observations are not quality-matched speedups. The internal expected-substring checks on short factual prompts are smoke tests, not a general language-model quality benchmark.}
\label{tab:moe}
\end{table}

\subsection{KV Mobility: Functional Validation}
\label{sec:eval_kv}

Internal KV validation reports describe hardware checks with two two-stage chains and KV coordination enabled: move a session from chain~A to chain~B, retain the prior holder, and compare the resumed continuation with a fresh-session cold prefill. The reported matrix covers \texttt{ov-runtime}, \texttt{ov-dist-spec}, \texttt{gemma4}, \texttt{qwen36-moe}, and an OpenVINO-IR sparse-MoE fixture. The certificate requires matching output, a cross-chain pull hit, evidence that the destination consumed the imported state rather than its own cache, downstream restoration, and absence of restore/shape failures. These are internal functional checks; their logs and harness are not public artifacts, and they are not new measurements from this paper's reproduction scripts.

Scope matters. The sparse-MoE cell in those checks uses a small fixture, not the full disk-streamed K2.6 workload of Table~\ref{tab:moe}. An internal Qwen3 QK-normalized dense-model check reported a warm/cold divergence, so the passing matrix is not a guarantee for an entire model family. Separate internal stream-resume reports cover a native sparse-MoE fixture and interrupted-stream checks for prefix preservation, continuation, a forced-prefix rescue event, and a clean terminal. Those tests establish a different contract from identical output to an uninterrupted run, and do not establish full-size-model throughput.

These checks establish compatible-prefix transfer and scoped continuation behavior. Their outcome is functional: state was consumed by the destination, the asserted output or prefix checks passed, and the run ended with the expected terminal behavior. They do not establish a latency advantage over cold recomputation or a durable session service. KV coordination and stream resumption retain independent, default-off controls and their engine/model restrictions.

\section{Comparison with Hyperconverged Enterprise AI Infrastructure}
\label{sec:comparison}

This section compares the configurations and listings documented in the June 2026 vendor-source snapshot, with reference links rechecked in September 2026: Cascadia's hardware does not overlap the incumbent platforms' supported configurations, so any head-to-head throughput number would conflate silicon with software. We compare the four platforms whose architectures are most fully documented (IBM, Nutanix, VMware, and HPE) and survey the wider landscape in \S\ref{sec:landscape}. Every vendor fact below cites the vendor's own documentation or a primary listing; Table~\ref{tab:comparison} summarizes.

\begin{table}[p]
\centering
\scriptsize
\setlength{\tabcolsep}{3pt}
\begin{tabular}{p{1.55cm}p{2.35cm}p{2.3cm}p{2.25cm}p{2.1cm}p{2.25cm}}
\toprule
Dimension & IBM (watsonx / OpenShift AI) & Nutanix (Enterprise AI) & VMware (Private AI Fdn.) & HPE (Private Cloud AI) & Cascadia \\
\midrule
Minimum footprint & OpenShift cluster (3 control-plane HA) + Software Hub + GPU workers; KServe; Serverless / Service Mesh in serverless mode~\cite{rhoaiinstall2026, rhoaiserving2026, redhatocp2026} & 3 control-plane + 3 worker K8s nodes~\cite{nutanixnaireq2025}; Cisco CVD entry: 4 HCI nodes, 2$\times$L40S each~\cite{ciscocvd2024} & Full VCF stack (SDDC Mgr., vCenter, NSX, Supervisor); $\ge$3 GPU-enabled hosts~\cite{vmwarepaif2026} & Turnkey rack; 3 dedicated control nodes + GreenLake cloud mgmt.\ plane~\cite{hpepcai2026} & $N\ge1$ AI PCs; no dedicated tier; CA off-path \\
\addlinespace
Accelerator floor & NVIDIA A100 / H100 / L40S class; RHEL AI min.\ L4 24\,GB; Gaudi 3 tech preview~\cite{rhelai2025hw} & NVIDIA-only list (L40S, A100, H100, H200)~\cite{nutanixnaifaq2026}; CPU mode: Xeon AMX, $\le$10B~\cite{nutanixamx2025} & vGPU-capable datacenter NVIDIA $\times$3 hosts; MIG unsupported with NIM~\cite{vmwarepaif2026} & G1: 4$\times$L40S small production config; 2$\times$H100 NVL developer system~\cite{hpepcai2026} & CPU + iGPU + NPU already in each AI PC \\
\addlinespace
Control plane & Kubernetes mandatory; operator stack; license metering in-cluster & Kubernetes mandatory (CNCF); Prism for HCI layer & VCF private-cloud control plane mandatory~\cite{vmwarepaif2026} & 3 control nodes/rack; mgmt.\ plane in HPE cloud~\cite{hpepcai2026} & No dedicated request scheduler; CA handles fleet management \\
\addlinespace
Scaling model & Replica serving via KServe~\cite{kserve2026}; distributed execution depends on runtime configuration & Replica serving via NIM/KServe; NIM also supports distributed execution~\cite{nimmultinode2026} & vLLM/NIM serving; deployment and networking are managed by VCF~\cite{vmwarepais2025} & Managed NIM serving on rack-scale GPU configurations~\cite{hpepcai2026} & Replicas and load-balanced chains; client-device pipeline mechanism~\cite{berenbaum2026shards}; disk-streamed MoE \\
\addlinespace
Licensing meter & Per-VPC on-prem; 67 VPC listed \$643,200/yr~\cite{awswatsonx2026}; per-accelerator add-on~\cite{rhaccel2026, redhatsubguide2026}; per-token/GPU-hr SaaS~\cite{ibmwatsonxpricing2026} & Per GB of aggregate GPU vRAM~\cite{nutanixlicensing2026, cdwnai2025} atop per-core platform licenses & Per-core VCF subscription + per-GPU NVIDIA AIE~\cite{vmwarepaif2026, broadcomexplore2025, broadcomvcfcores2026, nvidianvaie2026} & GreenLake subscription (3/5-yr terms) + per-GPU NVAIE~\cite{hpepcai2026, nvidianvaie2026} & Closed-source mesh; Apache-2.0 runtime; existing client hardware \\
\addlinespace
Air gap / edge & Disconnected OpenShift (heavyweight); Fusion HCI racks~\cite{ibmfusion2025} & Dark-site bundles; NKP Metal still K8s on servers~\cite{nkpmetal2026} & VCF 9.1 fleet ops incl.\ air-gapped / sovereign~\cite{broadcomvcf91} & G2 air-gapped configuration~\cite{hpepcai2026} & Static binaries + local models; disconnected is the default shape \\
\addlinespace
Trust \& audit & Platform boundary + logging & Platform boundary + logging & Platform boundary + logging & Platform boundary + logging & Per-response ed25519 receipts; hash-chained logs; $\le$60\,s revocation~(\S\ref{sec:trust}) \\
\bottomrule
\end{tabular}
\caption{Architectural comparison, vendor-documented facts only (June 2026 source snapshot; URLs rechecked September 2026). ``Accelerator floor'' refers to the smallest supported \emph{managed-inference} configuration, not to research possibilities.}
\label{tab:comparison}
\end{table}

\subsection{Footprint and Control Plane}

All four stacks require a functioning control plane before the first token is served: three control-plane nodes for highly-available OpenShift or for NAI's documented minimum, plus the serving operators (KServe, and in OpenShift AI's serverless mode Knative and Istio)~\cite{rhoaiinstall2026, rhoaiserving2026, nutanixnaireq2025, redhatocp2026}; the full VCF private-cloud stack with at least three GPU-enabled hosts for VMware~\cite{vmwarepaif2026}; and for HPE, three dedicated control nodes per production rack plus a management plane hosted in HPE's own cloud, an external dependency the on-prem pitch keeps except in the G2 air-gapped configuration~\cite{hpepcai2026}. That is infrastructure to operate, patch, and staff independent of any model served. Cascadia's equivalent is zero dedicated machines: scheduling and ingress are properties of every node, the CA's admission and fleet-management services are off the inference request path, and a one-node deployment runs the same software as a larger one.

\subsection{Hardware Gating and Capital Expenditure}

The incumbent floor is datacenter silicon: Nutanix's supported accelerator list is NVIDIA-only~\cite{nutanixnaifaq2026}, validated entry designs start at four nodes with dual L40S each~\cite{ciscocvd2024}, the smallest accelerator Red Hat supports for RHEL AI inference is an L4~\cite{rhelai2025hw}, VMware requires vGPU-capable datacenter GPUs across a minimum of three hosts~\cite{vmwarepaif2026}, and HPE's G1 small production configuration carries four L40S~\cite{hpepcai2026}. The CPU-only alternatives are server-class (Xeon AMX, up to roughly 10B parameters in Nutanix's own scoping~\cite{nutanixamx2025}). These configurations require suitable server hardware, which may be purchased or drawn from an existing deployment, in an industry context where enterprise GPU estates are reported to average single-digit utilization~\cite{venturebeat2026gpu}.

Cascadia's substrate is the AI PC fleet the organization already bought, so the capital expenditure for serving capacity is largely sunk and idle-cycle harvesting is the goal. We stop short of a TCO model, because utilization, power, and support costs vary too much by organization to claim a single number, but the difference in where the capital line begins is large.

\subsection{Scaling Model}

Model parallelism is available in both datacenter runtimes and Cascadia. NIM documents multi-node tensor- and pipeline-parallel deployment, including orchestration and communication requirements~\cite{nimmultinode2026}. The distinction here is how that execution is deployed: Cascadia composes advertised stages on admitted client machines and routes requests locally at each entry, while the compared products provide managed server-cluster environments. The companion paper supplies the 70B client-fleet runtime result~\cite{berenbaum2026shards}; this paper describes its integration with the mesh and reports the separate functional evidence in \S\ref{sec:chains}.

\subsection{Licensing Structure}

The incumbents meter by provisioned capacity: per virtual processor core for on-prem watsonx (with a public 67-VPC data point at \$643,200/year~\cite{awswatsonx2026}, plus a mandatory in-cluster license-metering service), per physical accelerator for Red Hat's AI add-on~\cite{rhaccel2026, redhatsubguide2026}, per gigabyte of aggregate GPU vRAM for NAI~\cite{nutanixlicensing2026, cdwnai2025} stacked on per-core platform licenses, per core for the VCF subscription beneath VMware's private-AI services with NVIDIA AI Enterprise per GPU on top~\cite{vmwarepaif2026, broadcomexplore2025, broadcomvcfcores2026, nvidianvaie2026}, and GreenLake subscription terms over HPE's rack configurations~\cite{hpepcai2026}. Capacity-metered pricing means the license bill scales with provisioned hardware rather than solely by completed requests. Cascadia's runtime is Apache-2.0 software, while its enterprise mesh is closed source. The hardware argument is the reuse of client machines; the runtime license does not establish the enterprise mesh's licensing terms or total cost.

\subsection{The Wider Landscape}
\label{sec:landscape}

The four platforms above are representative, not exhaustive. Cisco's AI PODs bundle UCS nodes (entry: a single L40S) under Intersight management with an optional OpenShift layer~\cite{ciscoaipods2025}. Dell's AI Factory and Lenovo's Hybrid AI Advantage wrap NVIDIA's reference architectures in validated designs and services rather than shipping their own serving control planes; Lenovo's scalable-unit design is itself evidence of the pattern, specifying two head nodes and three Kubernetes control-plane nodes of dedicated infrastructure before the first GPU serves~\cite{dellaifactory2026, lenovolp2311}. At the sovereignty pole, Google Distributed Cloud runs Gemini inside vendor-operated air-gapped racks~\cite{gdcairgap2026}, and Azure Local extends Arc-managed, per-core-metered Kubernetes to edge servers~\cite{azurelocal2026, azurelocalbilling2026}. All of these share the shape this paper contrasts: dedicated infrastructure, a control plane, datacenter accelerators. The closest any vendor comes to small-machine serving is NVIDIA's DGX Spark, whose documented reference configurations cover two units over a direct cable and four over a 200\,GbE switch~\cite{nvidiasparkstack2026}.

\subsection{Scope and Caveats}
\label{sec:caveats}

A few caveats scope the contrast. First, IBM has a real Intel story at the \emph{server} tier: IBM Cloud made Gaudi~3 available for production workloads in 2025~\cite{ibmgaudi2025}, and Granite~4.0's Nano variants run on laptops~\cite{ibmgranitenano2025}, but as model releases outside the managed platform; nothing in either vendor's portfolio manages inference across client-class Intel silicon. Second, Nutanix's AMX mode is a genuine non-GPU path, on server Xeons and scoped by Nutanix itself to models up to roughly 10B parameters~\cite{nutanixamx2025}. Third, the incumbents are legitimate at their design point, with thousand-user datacenter serving, mature multi-tenancy, enterprise support, and ecosystem certifications, and they are actively reducing their own frictions, as with Broadcom folding private-AI services into the base VCF subscription~\cite{broadcomexplore2025} and HPE's air-gapped variant~\cite{hpepcai2026}. Cascadia addresses a tier none of these products targets, the fleet of machines already on desks, and an organization can reasonably run both. Fourth, the serving measurements cover at most four nodes, with the one-to-three-node curve and four-node comparison reported as separate sessions.

\section{Limitations}
\label{sec:limitations}

The system assumes a single-operator, permissioned mesh. Tenant namespacing does not establish multi-organization federation, and CA-hosted admission and fleet management remain centralized operational dependencies outside the inference path. Node packages cover Windows and Ubuntu, with driver-dependent accelerator support. OVMS CPU/GPU workers expose continuous batching; the NPU path and embedded shard execution are sequential within an engine. Feasible-chain composition and capacity-based deployment use advertised resources, not a link-cost optimizer. The state plane subscribes to admitted peers, so its direct-stream design alone does not establish large-fleet scalability.

KV coordination and stream resumption are separately disabled by default and apply to compatible engine/model configurations. Incompatible state falls back to recomputation; unsupported stream resumption surfaces an error. A compromised admitted worker is inside the KV tenant trust boundary, snapshots are not a durable session service, and signed receipts establish provenance rather than numerical correctness. NPU non-determinism limits bytewise comparison across executors. NAT validation covers simulated network configurations. The performance results apply to the dated deployments and workloads reported here: whole-model response rates, paired latency observations, and runtime placement and MoE experiments. Chain transfer and continuation checks establish functional behavior rather than performance or availability guarantees. Single-session runtime observations and approximate expert reduction have the scope stated in their captions.

\section{Conclusion}
\label{sec:conclusion}

Cascadia composes ingress, request scheduling, and inference on the same client machines, with a certificate authority providing admission and fleet management outside the request path. Whole-model replicas and pipeline chains share an authenticated API, signed response receipts, and local audit logs. Session affinity and compatible KV transfer manage conversation state when routing changes; forced-prefix continuation has a separate contract. The June replica study records $3.10\times$ response throughput at three nodes relative to one, and a separate May four-node deployment records $4.06\times$ the throughput of direct single-node serving. The paired latency and runtime observations characterize the tested configurations, while functional checks establish the scope of chain and KV behavior. These results support client fleets as a practical serving substrate with a different deployment footprint from managed datacenter clusters. Benchmark scripts, curated records, and the evidence map accompany the paper in the \href{https://github.com/labscommunity/cascadia-architecture-paper}{public repository}.

\bibliographystyle{plainnat}
\bibliography{references}

\end{document}